\documentclass[oldversion,twocolumns]{aa}
\usepackage{graphicx}

\usepackage{txfonts}
\usepackage{natbib}
\bibpunct{(}{)}{;}{a}{}{,}
\def\na{\ion{Na}{I}}
\def\ca{\ion{Ca}{I}}

\def\co{CO\,(2--0)}
\def\kms{$\mbox{km s}^{-1}$}

\begin{document}

\title{Central K-band kinematics and line strength maps of NGC\,1399
\thanks{Based on observation collected at the ESO Paranal La Silla
Observatory, Chile, during the Science Verification of SINFONI.}}

 \author{Mariya Lyubenova
          \inst{1}
          \and
          Harald Kuntschner\inst{2}
	  \and
	  David R. Silva
	  \inst{3}
          }

 \institute{ESO, Karl-Schwarzschild-Str. 2, D-85748 Garching bei
 M\"unchen, Germany, \email{mlyubeno@eso.org}
 \and Space Telescope European Coordinating
 Facility, Karl-Schwarzschild-Str. 2, D-85748 Garching bei
 M\"unchen, Germany, \email{hkuntsch@eso.org}
 \and National Optical Astronomy Observatory, 950 North Cherry Ave., Tucson, AZ, 85748
 USA, \email{dsilva@tmt.org}}

\date{Received 21 December 2007/ Accepted 12 March 2008}

\abstract{In this paper we present for the first time high spatial
resolution K-band maps of the central kinematical and near-infrared
spectral properties of the giant cD galaxy in the Fornax cluster,
NGC\,1399. We confirm the presence of a central velocity dispersion
dip within $r \le 0\farcs2$ seen in previous long-slit studies. Our
velocity dispersion maps give evidence for a non-symmetric structure
in this central area by showing three $\sigma$ peaks to the
north-east, south-east and west of the galaxy centre. Additionally we
measure near-IR line strength indices at unprecedented spatial
resolution in NGC\,1399.  The most important features we observe in
our 2-dimensional line strength maps are drops in \na\/ and \co\/
line strength in the nuclear region of the galaxy, coinciding
spatially with the drop in $\sigma$.  The observed line strength and
velocity dispersion changes suggest a scenario where the centre of
NGC\,1399 harbours a dynamically cold subsystem with a distinct stellar
population.

}


\maketitle
\section{Introduction}

The centres of giant early-type galaxies are interesting laboratories
for exploring important questions about the physics of early-type
galaxy assembly and star formation history. In particular, the
ultimate goal is to tie the formation of global properties to such
central properties as nuclear stellar populations, super-massive black
holes, and active galactic nuclei.

The recently developed technique of near-infrared integral field
spectroscopy in conjunction with adaptive optics offers a new window
into these astrophysical questions by probing spatial scales
($\sim$0\farcs1 or better) heretofore only accessible to the
instruments on the Hubble Space Telescope (HST).  At the distance of
the Fornax cluster ($\sim$ 20 Mpc) this corresponds to a spatial
resolution of about 10\,pc, which can be achieved over a
$3\arcsec\times3\arcsec$ field-of-view with e.g.,
VLT/SINFONI. Although with limited field of view, this setup can be
used to scrutinise compact stellar subsystems such as globular
clusters or the nuclear regions of galaxies.

Today it is widely accepted that many massive and bright early-type
galaxies contain super-massive central black holes (SMBH) with masses
reaching up to $10^9$\,M$_\odot$, which tightly correlate with the
physical properties of the host galaxies
\citep{kr95,geb00,ferrmerr00}.  Recently, it was shown that the nuclei
of low- and intermediate luminosity early-type galaxies in the Fornax
and Virgo clusters exhibit a (often blue) central ($<$0.02 effective
radii) light excess in contrast with high-luminosity galaxies
($M_B\,\la\,-20$), which usually show a flat luminosity profile (core)
at the same spatial scales \citep{cote06,cote07}. The mass fraction
contributed by the central light excess is similar to the relative
contribution of SMBH in giant galaxies, which could suggest a common
formation path \citep{wh06,cote06}. The nature of these central light
anomalies is still uncertain since until recently their spatial scales
were typically reachable only by HST imaging and in a few cases
HST/STIS spectroscopy.

In this paper we present for the first time high spatial resolution
two-dimensional maps of the velocity profile and the distribution of
two near-IR spectral features in NGC\,1399 -- the central giant
elliptical galaxy in the Fornax cluster ($M_B\,\la\,-21.08$). This
galaxy is known to harbour a SMBH \citep{sagl00,h06,geb07}, weak
nuclear activity \citep{ocn05} and shows a classical core in the
luminosity profile \citep{cote07}.

The paper is organised as follows: in Section 2 we present the
observations and data reduction steps.  Section 3 is devoted to the
analysis of the kinematics in NGC\,1399, while Section 4 presents the
measurements of near-IR line strength indices. The main observed
features and potential explanations are discussed in Section 5. Our
concluding remarks are given in Section 6.

\section{Observations and data reduction}
\subsection{Observations}
The observations were made with SINFONI - a near-infrared Integral
Field Unit (IFU) spectrograph, mounted in the Cassegrain focus of Unit
Telescope 4 (Yepun) of VLT at Paranal La Silla Observatory
\citep{eis03, bonnet04}. The instrument's field of view on the sky is
divided into 32 slices. The pre-slit optics allow us to chose between
three different spatial samplings: 0\farcs25, 0\farcs1 or 0\farcs025,
yielding a total field of view of 8$\arcsec\,\times\,8\arcsec$,
3$\arcsec\,\times\,3\arcsec$ or 0\farcs8\,$\times$\,0\farcs8,
respectively. On the detector each slitlet is imaged over 64 pixels in
a brick-wall pattern and thus each one is well separated from the
neighbouring slitlets on the sky. The Science Verification detector
used during our observations had a blemish affecting wavelengths
between 2.37 and 2.40\,$\mu$m of the spectrum from slitlet number
17. This defect had no impact on our investigation since it lay
outside of the wavelength region of interest.

%
%
\begin{center}
\begin{table}[htdp]
\caption{\label{tab:observations}NGC\,1399 observing log.}
\begin{tabular}{c c c c c }
\hline 
\hline 
UT date & t (s) & O+S Frames
& Total t (s) & OB ID\\ 
(1) & (2) & (3) & (4) & (5) \\
\hline
2004 Oct\, 05 & 2x300 & 4+4 & 2400 & A \\
2004 Nov 25 & 2x300 & 4+4 & 2400 & B \\
2004 Nov 26 & 2x300 & 4+4 & 2400 & C \\
2004 Nov 26 & 2x300 & 4+4 & 2400 & D \\ 
2004 Nov 26 & 2x300 & 4+4 & 2400 & E \\ 
\hline
\hline
\end{tabular}

\smallskip 
Notes: (1) Date of
observation, (2) exposure time per object pointing, (3)
number of Object + Sky frames, (4) total on source exposure time 
per observing block, (5) observing block ID.
\end{table}
\end{center}

Our study makes use of the 0\farcs1 scale with a grism covering the
K-band from 1.95 to 2.45 $\mu$m at a dispersion of 2.45 \AA/pix. In
this mode, SINFONI delivers a spatial sampling of $0\farcs05 \times
0\farcs1$. However, during data processing the observations were
re-sampled to $0\farcs05 \times 0\farcs05$ as allowed by our dithering
strategy. The spectral resolution around the centre of the band is
R$\simeq$3800, measured from arc lamp frames. NGC\,1399 was observed
on 5 October, 25 November, and 26 November 2004 as part of the SINFONI
Science Verification program. Observing conditions were generally good
with free-air optical seeing measurements ranging from 0\farcs4 to
0\farcs7, except in the night of October $5^{th}$, where they reached
1$\arcsec$.

Observing in the near-IR spectral domain implies several
challenges. One of them is the very bright night sky consisting of
many strong emission lines.  To remove the night sky signature we used
the standard near-IR technique of consecutively taking object (O) and
sky (S) frames.  Our target was observed in five observing blocks
(OBs) each containing a OSSOOSSO sequence (see
Table~\ref{tab:observations}).  Each individual object pointing had
two 300\,s integrations, the sky pointings only one. The various on
source pointings were dithered by 0\farcs05 to reject bad pixels and
assure better spatial sampling. In the course of the data reduction we
decided to exclude OB A, because of highly variable sky conditions
resulting in a poor sky subtraction.

In order to improve upon natural seeing the observations were carried
out using adaptive optics (AO) in natural guide star assisted mode. We
used the same guide star (2MASS\,J03382914-3526442) as in the study of
\citet{h06} which is located 17\farcs6 to the North of the galaxy
centre.

Together with the science observations appropriate telluric standard
stars were observed, as well as four velocity template stars covering
the spectral range K4III to K7III (see Table~\ref{tab:stars}).

%
%
\begin{center}
\begin{table}[htdp]
\caption{\label{tab:stars}List of velocity template stars. }
\begin{tabular} {llc}
\hline 
\hline 
Name & spectral type  & Source \\
(1) & (2) & (3) \\
\hline
HD 0005425 & K4III  & this paper \\
HD 0025211 & K4/5III  & this paper\\
BD -15 1319 & K5III  & this paper \\
BD -16 1418 & K7III & this paper\\
2MASS\,J20411845+0016280 & M0III  & 075.B-0495(A)\\
HD 0141665 & M5III & 075.B-0490(A)\\
2MASS\,J17093801-2718559& M5II-III & 075.B-0490(A)\\
\hline
\hline
\end{tabular}

\smallskip
Notes: (1) star name, (2) spectral type, (3) origin
of data.
\end{table}
\end{center}

\subsection{Basic data reduction and telluric correction}
\label{sec:reduction}

The observations were reduced with the ESO SINFONI Pipeline v. 1.6
\citep{modigli07}. Calibration files such as distortion maps, flat
fields and bad pixel maps, were obtained with the help of the relevant
pipeline tasks (``recipes'').  Finally, for each data set contained in
one OB the {\tt jitter} recipe extracts the raw data, applies
distortion, flat-field corrections, wavelength calibration and stores
the combined sky-subtracted spectra in a 3-dimensional data cube.  The
same pipeline steps were also used to reduce the observations of the
telluric and velocity template stars. The {\tt jitter} recipe also
provides a one-dimensional spectrum extracted from an optimal
aperture, which worked well for our star observations.

The accuracy of the wavelength calibration was checked by comparing
the position of known sky emission lines to the wavelength calibrated
sky spectra. Where needed small shifts of up to one pixel ($=
2.45$\,\AA) were applied to our galaxy observations.

The next step is to correct the observations for telluric absorption
lines, which are especially deep at the blue end of the K-band. These
spectral lines may vary with time and position on the sky.  For each
observing block, one solar-type telluric standard star at a similar
air-mass was observed. Each telluric star was divided by a scaled and
Gaussian broadened solar spectrum to match the resolution of our
observations. In this way we removed the spectral features typical for
solar type stars as well as ensuring a relative flux calibration.  The
last step before applying the telluric correction was to scale and
shift the telluric spectrum for each individual data cube by a small
amount ($< $ 0.5 pix) to minimise the residuals of the telluric lines
(for details of the procedure see Silva, Kuntschner \& Lyubenova
2008).  Finally, each data cube for the galaxy and velocity template
stars was divided by the optimised telluric spectrum to yield fully
calibrated data.

In order to achieve the best signal-to-noise ratio we combined four
individual data cubes (OB ID: B, C, D, E). For this purpose we
identified the centre of the galaxy in each data cube with an accuracy
of one pixel ($= 0.05$\arcsec) and re-centred them. The final
combination was performed using a sigma-clipping pixel reject
algorithm.

The next step was to spatially bin the final data cube to a roughly
constant signal-to-noise ratio which allows us to measure reliable
kinematics and line strengths. This was done using the adaptive
binning method of \citet{cc03}. Using our velocity template stars we
first performed Monte Carlo simulations to determine the minimum S/N
needed to recover the higher order moments $h_{3}$ and $h_{4}$ of the
line-of-sight velocity distribution (LOSVD) to a precision of about
$\pm$\,0.04. This also means that the recession velocity and velocity
dispersion can be recovered with an error of about
$\pm$\,17\,\kms\/. This accuracy is comparable with or better than
what was achieved in previous studies, while we also preserve a good
spatial resolution. For our data set and the spectral region 2.2 -
2.3\,$\mu$m we found that a S/N of about 100 is needed (see also
Section~\ref{sec:kinematics}). The resulting bins have a typical size
of $\sim$ 0\farcs15 (5-6 spaxels combined) in the central regions, and
approximately 0\farcs2 ($>$10 spaxels combined) for radii
$>$1\arcsec. Bins which contain more than 30 spaxels (a few bins at
the edges of the field of view) are excluded from the analysis,
because of higher systematic effects, and are marked with black colour
on the kinematical and line strength maps. During the binning
procedure we derive pseudo error spectra by determining the standard
deviation at each wavelength. While this procedure is not fully
correct since not all spaxels in a given bin are independent (e.g.,
seeing effects) it takes into account other error sources such as
sky-subtraction and telluric correction residuals.

\subsection{Effective spatial resolution}
\label{sec:seeing}

A key parameter to characterise our data is the effective spatial
resolution, which was achieved in the combined data cube before
binning.  Direct measurements of the effective point spread function
were not obtained during the nights of our observations. Therefore, we
estimate the effective seeing in two other ways.

\begin{figure}
\resizebox{\hsize}{!}{\includegraphics[angle=0]{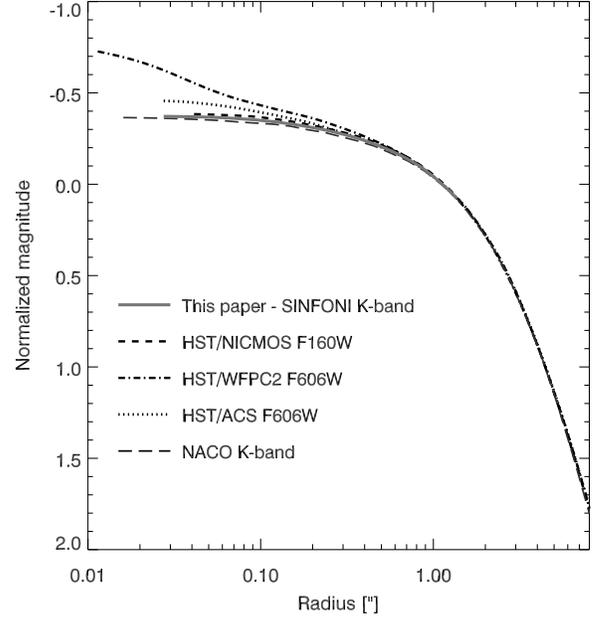}}
\caption{\label{fig:1399lum} Normalised luminosity profile of
NGC\,1399 as reconstructed from our data cube (solid grey line). As a
comparison we show HST/NICMOS F160W (short dashed line) and HST/ACS
F606W (dotted line) observations (retrieved from the HST Archive) as
well as HST/WFPC2 F606W (dash-dot line) profile from \citet{geb07} and
a VLT/NACO K-band profile from Houghton (private communication,
program 078.B-0806, PI Houghton) -- long dashed line. Profiles are
normalised in the region 1\arcsec\,$\le\,r\,\le$\,1\farcs4. The WFPC2
light profile is super-sampled and deconvolved for the instrumental
PSF which at least partly explains the differences between the WFPC2
and ACS data, obtained in the same filter (F606W).}
\end{figure}

First, we compared the reconstructed and radially averaged luminosity
profile from our data cube with the results from a HST/NICMOS image
(GO 7453, PI Tonry; camera NIC2, sampled at 0\farcs0755 per pixel). We
found that the two luminosity profiles agree very well even without
any further broadening of the NICMOS data suggesting roughly similar
PSF sizes with a FWHM~$\simeq 0\farcs17$ (see Fig.~\ref{fig:1399lum}).

Secondly, we independently confirmed the above estimate from our own
data cube by measuring the FWHM of a point source like object
(presumably a globular cluster) located at about 1\farcs45 south-east
from the NGC\,1399 nucleus. Globular clusters at the distance of the
Fornax cluster might be extended. However, here we assume it to be a
point source and thus obtain a conservative estimate for the seeing. A
simple Gaussian fit yields a FWHM~$= 0\farcs11$, very much in
agreement with the results from the comparison of the luminosity
profiles.

We conclude that the effective spatial resolution of our data can be
described with having a FWHM of about 0\farcs15.

%
%

\section{Kinematics}

In this section we present our kinematical analysis of the central
region of NGC\,1399. Our K-band IFU spectroscopy is used to probe the
velocity profile of the galaxy's centre to HST like spatial
resolution.

\subsection{Extraction of the kinematics and comparison with earlier studies}
\label{sec:kinematics}

The extraction of the kinematical information from the final galaxy
data cube was performed with the help of the Penalised Pixel-Fitting
method (pPXF) developed by \citet{ce04}.  The success of this method
relies on the provision of a good set of template spectra, which match
the galaxy spectra as well as possible. We found that the template
stars observed in conjunction with the NGC 1399 observations (covering
K4III to K7III) did not yield fully satisfactory fits and thus we
searched the VLT/SINFONI archive for further template stars
particularly covering later spectral types. We found three suitable
stars with spectral type M0III, M5III and M5II-III. Fully reduced
spectra were kindly provided by M. Cappellari (private
communication). In our final analysis seven velocity template stars
(see Table~\ref{tab:stars}) were used in the fitting procedure to
determine the recession velocity, velocity dispersion and the $h_3$
and $h_4$ Gauss-Hermit coefficients for each bin in the data cube.

\begin{figure}
\resizebox{\hsize}{!}{\includegraphics[angle=0]{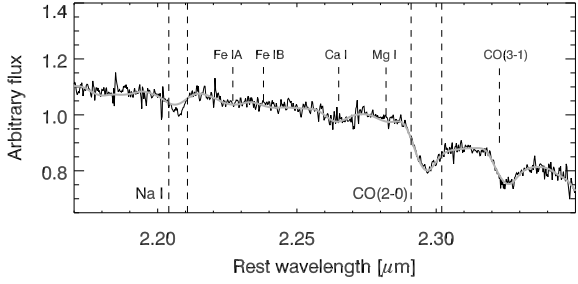}}
\caption{\label{fig:1399spec}Normalised spectrum of NGC\,1399 taken
from the central bin of our data. The grey line shows the best fitting
composite template as derived in our kinematics measurements. The
vertical dashed lines show the extent of the central bandpass of the
\na\/ and \co\/ indices. The positions of several other absorption
features are indicated.}
\end{figure}

The fitting was performed in the spectral range between 2.18\,$\mu$m
and 2.35\,$\mu$m where several prominent absorption features allow
fitting (see Fig.~\ref{fig:1399spec}). However, we excluded the region
2.20\,$\mu$m-- 2.22\,$\mu$m (around the \na\/ feature), because none
of the template stars was able to match the strength of the \na\/
absorption in NGC1399. This is not surprising, since \citet{silva2008}
showed that the \na\/ absorption strength is typically stronger in
early-type galaxies compared with Galactic open cluster stars. In the
fitting procedure typically only 3-4 stars were used to achieve a good
fit over the full wavelength extent. The stars which received the
highest weighting were of spectral type M0III and M5III.

%
%

We derive one-sigma error estimates from the pPXF code, based on our
error spectra (see Sect.~\ref{sec:reduction}).  The validity of our
error spectra is confirmed since the reduced $\chi^{2}$ values derived
by the code were typically close to unity.  Furthermore, we performed
an empirical error estimation by assuming spherical symmetry in the
galaxy and deriving the standard deviation for v, $\sigma$, $h_3$ and
$h_4$ measurements in a ring of radius 0\farcs8 to 1\arcsec.  The
errors derived with this procedure agree well with the pPXF internal
error estimates and are 17.2\,\kms, 19.4\,\kms, 0.03 and 0.05 for v,
$\sigma$, $h_3$ and $h_4$, respectively.

The kinematics of NGC\,1399 have been studied several times in the
past using long-slit spectroscopy, mainly for the purpose of
determining the mass of the super-massive black hole in its
centre. Velocity dispersions as high as $\sim$\,450\,\kms\/ make this
galaxy an extreme and very interesting case in the $\sigma$-black hole
mass relation \citep{ferrmerr00,geb00}.

\citet{sagl00} used ground based optical observations and reported
little rotation ($\leq$ 30\,\kms\/) and compatibility with the
presence of a central black hole of
M\,$\approx$\,5\,$\times$\,10$^8$\,M$_\odot$.  \citet{h06} studied
NGC\,1399 with ground based adaptive optics assisted observations in
the near-IR using the ESO VLT instrument NACO. Their analysis reveals
a kinematically decoupled core and double-peaked velocity dispersion
across the centre. The modelling requires a central black hole with
M\,$\approx$\,1.2\,$\times$\,10$^9$\,M$_\odot$ and a strongly
tangentially biased orbit distribution.  The most recent study is the
one by \citet{geb07}. They analysed HST/WFPC2 optical imaging and
HST/STIS spectroscopy in combination with the \citet{sagl00}
data. They see a dramatic increase in the velocity dispersion at about
0\farcs5 on both sides of the galaxy and a central drop, which they
interpret as tangentially biased orbital distribution. Their best fit
model requires a black hole with mass
$(5.1\pm0.7)\,\times$\,10$^8$\,M$_\odot$.
%
%
\begin{figure}
\resizebox{\hsize}{!}{\includegraphics[angle=0]{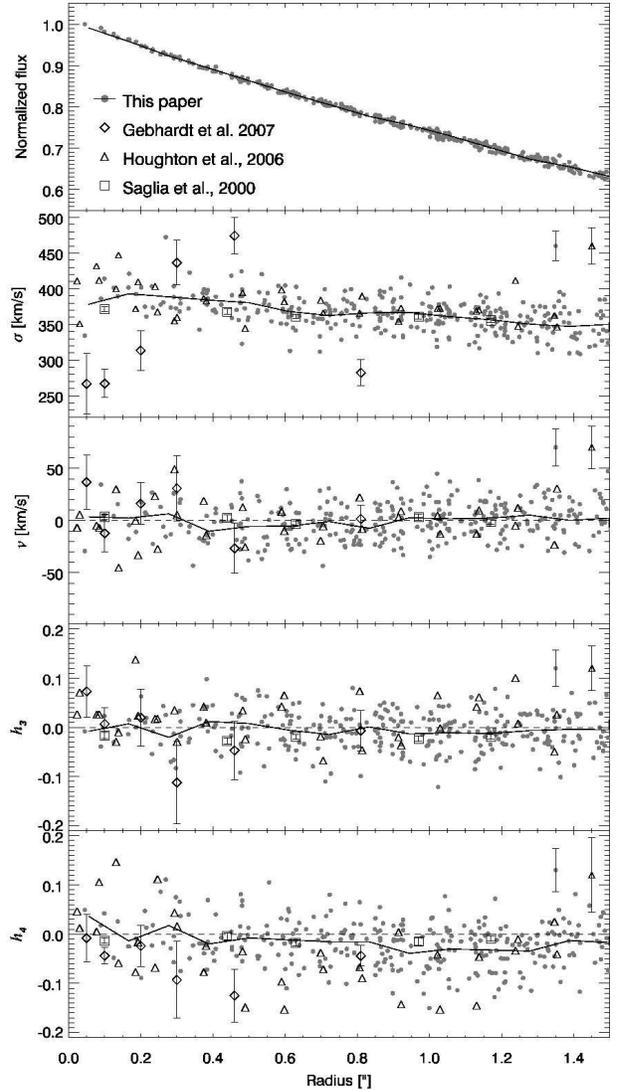}}
\caption {\label{fig:1399sig_comp} Light distribution and the first
 four moments of the velocity profile for NGC\,1399. Top plot:
 Luminosity profile, as inferred from our SINFONI data. The solid line
 represents the mean luminosity in 0\farcs1 bins. Bottom panels:
 radial profile of the kinematics. SINFONI observations are marked
 with solid grey circles. Empirical error estimates (for details see
 Section~\ref{sec:kinematics}) are shown with an error-bar and a solid
 symbol in the top-right corner of each plot. The solid line
 represents the mean value of each moment in 0\farcs1 bins. Results
 from earlier studies are shown with different symbols as explained on
 the top plot. Representative errors from \citet{h06} are shown with
 an error bar in the top-right corner of each panel with the triangle
 symbol. Errors from \citet{sagl00} and \citet{geb07} are plotted with
 the data points.}
\end{figure}

In Fig.~\ref{fig:1399sig_comp} we show our kinematical measurements
together with data from previous studies. We use radial plots for this
first-order comparison, since the observations in the earlier studies
were obtained with long-slits at different position angles (110\degr\/
in \citet{sagl00}, 117\degr\/ in \citet{geb07} and 5.06\degr\/ in
\citet{h06}). We plot data from the full two-dimensional region
covered by our observations.

For radii $>$\,0\farcs3 we find in general good agreement between our
measurements and previously published results. However, we cannot
confirm the large negative $h_4$ values seen by \citet{h06} although
we also measure negative $h_4$ values at these radii. We also cannot
confirm the velocity dispersion measurements of \citet{geb07} for
their two outermost data points.

For radii $<$\,0\farcs2 our measurements appear to be between the
measurements of \citet{h06} and \citet{geb07}. Although the studies do
not agree well for the inner velocity dispersion gradient, all three
of them show a pronounced drop in the very centre. In this comparison
the HST/STIS data probably have the best spatial resolution ($\simeq
0\farcs05$; \citet{geb07}), however, even accounting for seeing
differences it is difficult to reconcile the data sets.  Similarly, we
cannot find any evidence in our data for the extreme, positive $h_4$
values seen in the \citet{h06} study.

\subsection{Kinematical maps}
\label{sec:sigma_maps}

\begin{figure*}
\resizebox{\hsize}{!}{\includegraphics[angle=0]{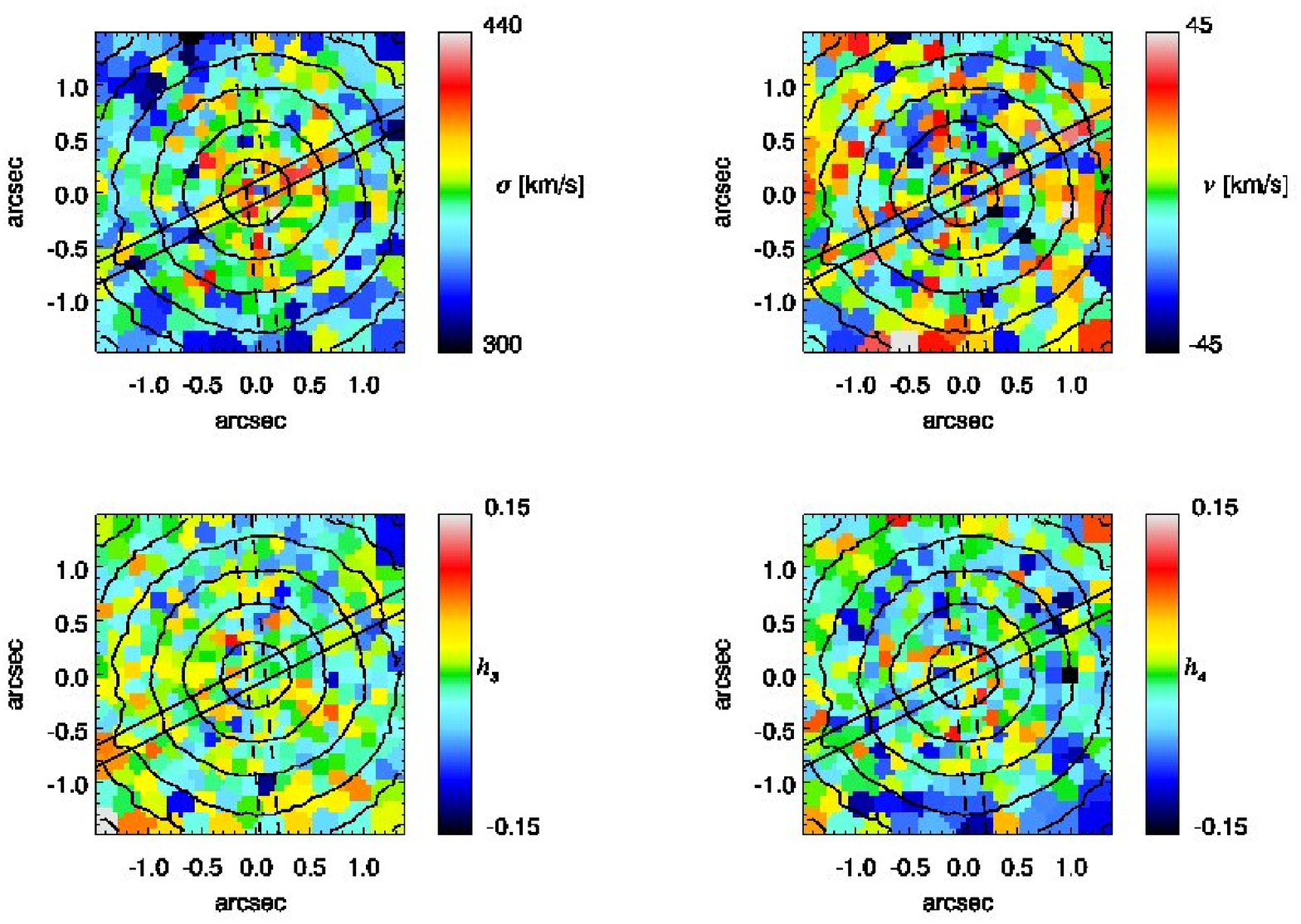}}
\caption {\label{fig:1399kin} Maps of the first four moments of the
velocity profile in the central 3$\arcsec$ of NGC\,1399. Top left -
velocity dispersion, top right - velocity. Bottom left - $h_3$ and
bottom right - $h_4$. North is up, east - to the left. Slit
orientation and width of the \citet{geb07} study are shown with solid
lines (PA\,=\,117\degr, width\,=\,0\farcs2), the slit of \citet{h06}
is shown with dashed lines (PA\,=\,5.06\degr, width\,=\,0\farcs172).
Overplotted are levels with constant surface brightness, as derived
from our data.}
\end{figure*}

Given that we have full two-dimensional kinematic maps, we can explore
if there is a more complex kinematical structure in the centre of
NGC\,1399, which might explain the measurement differences reported
above.  In Fig.~\ref{fig:1399kin} we present our maps of $\sigma$,
$v$, $h_3$ and $h_4$. We also overplot the slit orientations and
widths of \citet{h06} and \citet{geb07} studies for better
illustration. The \citet{sagl00} slit covers almost our full
field of view, thus it is not shown in Fig.~\ref{fig:1399kin}

The velocity dispersion map shows rising values towards the centre
with a distinct drop in the middle. However, the inner
$\sim$\,0\farcs5 appear to be asymmetric, and show three peaks to the
north-east, south-east and west of the centre. Fig.~\ref{fig:1399kin}
shows that the \citet{geb07} slit intersects two of the high-$\sigma$
features, which can potentially explain the disagreement with
\citet{h06}, whose slit does not cover any of them. If confirmed, the
velocity dispersion structures presented in this work may imply the
need for more detailed dynamical modelling for the nuclear parts of
the galaxy.

Our velocity map does not show any significant ordered motion within
our errors ($\simeq$17\,\kms).  The $h_3$ map is consistent with a
value of zero and thus no structure, while the $h_4$ map is mildly
positive in the very centre and mostly negative for radii
$>$\,0\farcs4.

%
%

\section{Near-IR line strengths}
\label{sec:indices}
In this section we discuss the two-dimensional distribution of two
near-IR absorption features -- \na\/ close to 2.2\,$\mu$m and \co\/ at
2.3\,$\mu$m (see Fig.~\ref{fig:1399spec}) -- at a spatial resolution
and scale only accessible by HST observations in the past.

\subsection{Index definitions and measurements}
\label{sec:definitions}
Index definitions for \na\/ and \co\/ were taken from
\citet{frog01}. The line strength of \na\/ is measured in \AA,
identical to the method used by the optical indices in the Lick/IDS
system \citep{wort94}, where the central bandpass is flanked to the
blue and red by pseudo-continuum band passes. \co\/ is measured in a
similar way but due to the lack of a well defined continuum on the red
side of the CO feature, \citet{frog01} define four pseudo-continuum
band passes on the blue side. Thus the continuum used to derive the
value of the \co\/ index is an extrapolation from the blue wavelength
region.

The observed spectrum of a galaxy is the convolution of the integrated
spectrum of its stellar population(s) by the instrumental broadening
and line-of-sight velocity dispersion (LOSVD) of the stars. These
effects broaden the spectral feature and in general reduce the
observed line strength compared to the intrinsic value. In order to
compare our observations with other studies, often obtained with
different instruments, and eventually with population synthesis
models, one needs to calibrate the data to a common system. Since an
averaged spectrum of the observations reported in this study was
already used in the \citet{silva2008} VLT/ISAAC study of early-type
galaxies, we decided to use that spectral resolution here as
well. Therefore, all galaxy spectra were broadened to a spectral
resolution of 6.9\,\AA\/ (FWHM) matching the resolution of ISAAC in
the K-band.

Using our measurements of the velocity for each bin (see
Sect.~\ref{sec:kinematics}) we can determine line strength
values. However, before we can analyse the line strength maps we need
to account for line broadening due to the internal motion of stars in
the galaxy. In order to compare the index measurements for regions
with different LOSVD or between different galaxies, we calibrate the
indices to a zero velocity dispersion. Our LOSVD corrections were
derived by broadening the velocity template stars to velocity
dispersions ranging up to 440\,\kms and $h_3$, $h_4$ values between
$-0.20$ and $+0.20$. For each index we parametrised the LOSVD
corrections following the method by \citet{kun04}. Using our own
kinematical measurements we calculated for each bin and index a
correction factor which is applied to the raw line strength
measurements. At $\sigma = 400$\,\kms\/ ($h_3$, $h_4\,=\,0.0$) the
correction factor is 1.53 and 1.16 for \na\/ and \co,
respectively. The typical error on the correction factor as derived
from the differences between the template stars is 3.3\% for \na\/ and
1.3\% for \co.

Using our error spectra we compute index errors via Monte Carlo
simulations, where we take into account photon noise, recession
velocity errors, and LOSVD correction errors.  Following the procedure
applied for the kinematics (see Sect.~\ref{sec:kinematics}) we also
derive empirical error estimates by computing the standard deviation
of index values in a region with radius between 0\farcs8 and 1\arcsec.
For the \na\/ index, the empirical estimate is 0.5\,\AA, and thus
slightly higher than our internal estimate of 0.38\,\AA. For the \co\/
index we find a much larger difference with an empirical error of
0.6\,\AA\/ as compared to 0.17\,\AA\/ for the internal value. Line
strength indices can be very sensitive to continuum shape effects,
especially if they cover large wavelength ranges as is the case for
the \co\/ index.  We consider this the main reason for increased
errors \citep[see also][]{silva2008}. In the following we adopt the
empirical error estimates in our analysis.

\citet{silva2008} present a more complete set of spectral features in
the K-band, including the \ca\/ and two Fe indices (see
Fig.~\ref{fig:1399spec}). Here we will not discuss these additional,
weak features, because our chosen SNR per bin in combination with the
very high velocity broadening of NGC\,1399 makes them very difficult
to measure.

\subsection{Line strength maps}
\label{sec:index_maps}
\begin{figure*}
\resizebox{\hsize}{!}{\includegraphics[angle=0]{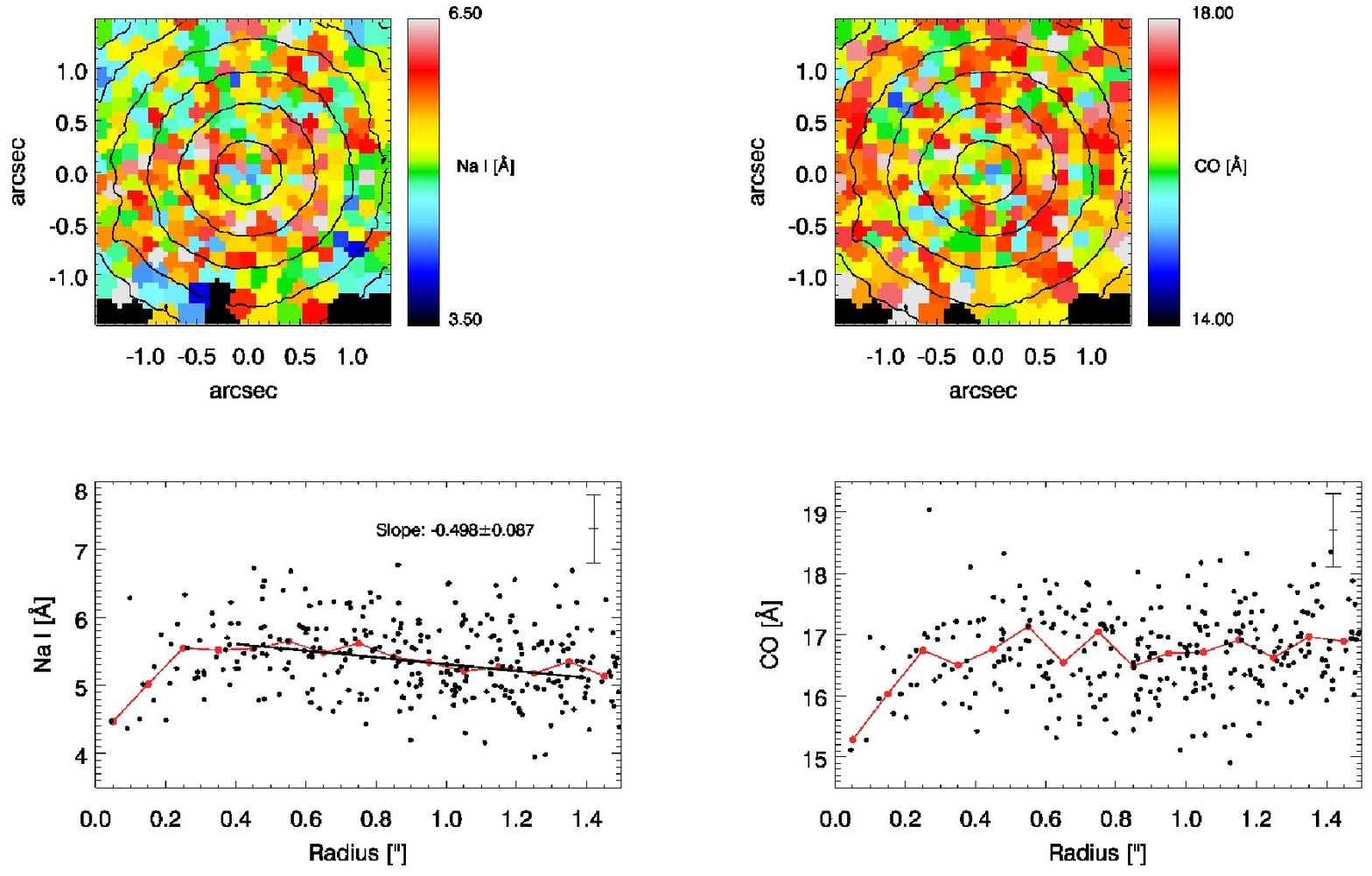}}
\caption {\label{fig:index} Top plots: Maps of \na\/ and \co\/ in the
central 3\arcsec of NGC\,1399.  Overplotted are levels with constant
surface brightness, as derived from our data. Bottom plots: Radial
distribution of the same indices. The red lines represent the median
index values in 0\farcs1 bins. The black line shows a linear fit to
the index values of \na\/ in the region
0\farcs4\,$<\,r\,<\,$1\farcs4. The slope value is given in the
plot. Typical index errors (see Sect.~\ref{sec:definitions}) are shown
with error bars in the upper right corner of the plots.}

\end{figure*}

In Fig.~\ref{fig:index} we show the distribution of \na\/ and \co\/
over the central $3\arcsec \times 3\arcsec$ of NGC\,1399.  In the
central 0\farcs2 we see a well pronounced drop of about 1\,\AA\/ and
1.5\,\AA\/ in \na\/ and \co, respectively. Below the maps we show the
radial distribution of both indices for better illustration. The red
line shows the median value of each index in 0\farcs1 wide bins. We
note, that \citet{h06} also found some evidence for a central CO drop
in their data, although their measurements were not
conclusive. Outside the nuclear region we find evidence for line
strength gradient in \na. The index decreases with radius with a slope
of $-0.498\pm0.087$ within 0\farcs4 and 1\farcs4 (the Spearman
correlation coefficient is $-0.21$ with a probability of 0.09\%
that there is no correlation). The linear fit to the data is
overplotted in Fig. ~\ref{fig:index}. For \co\/ we do not find a
significant line strength gradient over the observed radial range.

%
%

\section{Discussion of the central velocity dispersion and line strength drops}

In this section we will discuss the significance of the central
$\sigma$ and line strength drops we described in
Section~\ref{sec:sigma_maps} and ~\ref{sec:index_maps} and explore a
few possible scenarios that could explain their origin.

\subsection{Technical aspects}
Firstly, we investigate the statistical significance of the central
drops in absorption strength of the \na\/ and \co\/ indices. Using a
two-dimensional Kolmogorov-Smirnov test we can reject the
null-hypothesis that the innermost five data points ($r < 0\farcs15$)
are drawn from the same distribution as the data points between
0\farcs3 and 1\arcsec\/ with a significance level of 3.4\% and 0.4\%
for the \na\/ and \co\/ indices, respectively. While for the \na\/
index the significance is only marginal, both index drops taken
together are believable. Another test we performed is to fit a simple
two component model, where we fit the data with two linear functions
with a break radius at $r=0\farcs25$. The steep slopes measured for
the central component are different from zero at four and nine sigma
level for the \na\/ and \co\/ indices, respectively.

Secondly, we explore the possibility that the line strength drops
could be caused by wrong kinematics (e.g., too low velocity
dispersion) which would lead to wrong (e.g., too low) LOSVD
corrections. Due to the large velocity broadening seen in the centre
of NGC\,1399 we have to apply significant LOSVD corrections (see
Sect.~\ref{sec:indices}). However, even if we assume the extreme case
of a LOSVD ($\sigma = 370$\,\kms, $h_3=0.0$ and $h_4=0.0$) constant
with radius we still find highly significant drops for both indices.

In summary we find that the line strength drops are significant at
least at the 3-$\sigma$ level and are unlikely to be produced by
technical/calibration problems or by a chance statistical
effect. Furthermore the observed features in the velocity dispersion
and line strengths are present also when one performs the same
analysis with the individual data cubes before combination, although
the noise level raises.

\subsection{Astrophysical explanations}

 Central velocity dispersion drops have so far been observed mainly in
  spiral galaxies \citep[e.g.][]{bottema97,emsellem01,marquez03,fb06}.
  $\sigma$-drops have also been found in a few early type galaxies
  \citep[e.g.][] {graham98, simien02, pinkney03, emsellem04}. For
  spiral galaxies \citet{wozniak03} suggested that these
  $\sigma$-drops are the result of gas accretion followed by star
  formation and are therefore indicative of a recent in-fall of
  dissipative material. In a follow up study \citet{wozniak06} showed
  that the life time of such $\sigma$-drops can be long
  ($\ge$\,1\,Gyr), however this requires continuous star formation in
  the nuclear region at least on a level of
  $1\,M_{\odot}\,yr^{-1}$. Without continued star-formation, the
  amplitude of the $\sigma$-drops decreases rapidly.

  Velocity dispersion drops were found in about half of the sample of
  spiral galaxies, studied by \citet{peletier07}. Their stellar
  populations analysis, however, shows that the ages of these inner
  components are not significantly younger than the centres of spiral
  galaxies without $\sigma$-drops as would be expected from the above
  scenario. This may indicate that the central, dynamically cold
  components can be more long lived.

  Typically the observed $\sigma$-drops in spiral galaxies have
  spatial scales of a few hundreds parsecs. For NGC\,1399 such a large
  scale $\sigma$-drop is not observed \citep{graham98}. The
  $\sigma$-drop we observe in NGC\,1399 has a diameter of only
  $\sim$30\,pc (corresponding to 0\farcs3). Similar size
  $\sigma$-drops have been seen with HST/STIS spectroscopy in the
  elliptical galaxies NGC\,4649 and NGC\,4697 by \citet{pinkney03}
  indicating that these small scale drops may be a more wide spread
  phenomenon in early-type galaxies.

On the basis of their dynamical modelling \citet{geb07} state that
NGC\,1399 possesses extremely tangentially biased orbital distribution
for radii between 0\farcs1 and 0\farcs5 which is responsible for the
$\sigma$-drop, consistent with the findings of \citet{h06}. Both
studies explore various possibilities for the progenitor of these
extreme orbits, ranging from nuclear activity, an eccentric disc, a
torus to a stellar cluster having fallen in on a purely radial orbit.

We can empirically test the hypothesis for the presence of a low
velocity dispersion subsystem in the nucleus of NGC\,1399 causing the
$\sigma$-drop. We performed simple simulations by combining an
averaged galaxy spectrum taken at radii between 0\farcs2 and 0\farcs3
($\sigma\sim 400$\,\kms\/) with a velocity template star, broadened to
velocity dispersions from 15 to 150\,\kms\/ and a range of luminosity
weightings. We can reproduce our measurements for the inner 0\farcs15
of NGC\,1399 ($\sigma\,<$\,360\,\kms\/ and positive $h_4$) with
$\sim$15\% of the light originating from a central stellar subsystem
with $\sigma$ up to 100\,\kms. Using the surface brightness profile of
NGC\,1399 \citep[e.g. from][]{geb07} we can estimate the surface
brightness of such a subsystem. We find that a central, low velocity
dispersion ($\sigma \le 100$\,\kms) stellar subsystem with $V \simeq
20.3^m$ is able to explain the central drop in velocity dispersion for
NGC\,1399. Such a V-band luminosity is comparable to the brightest
globular clusters in NGC\,1399 or bright nuclei of dwarf galaxies in
the Fornax cluster \citep[e.g.][]{dirsch03,miske06}.

If such a subsystem is present in the nucleus, we can expect that it
will not have the same stellar population properties as the underlying
galaxy since it might have been formed in different time scales, under
different physical conditions or in a different environment. Thus we
can expect that it will have some influence on the line strengths in
the same region. Indeed, as shown in Sect.~\ref{sec:index_maps}, we
find line strength changes with the same extent as the $\sigma$-drop.
Nevertheless, index changes might be due to different reasons than
stellar population effects. A plausible explanation for the reduced
line strengths of \na\/ and \co\/ in the central regions could be the
presence of nuclear activity causing the continuum level to be raised.
\citet{oliva99} show that the nuclear continuum in the near-IR
spectral domain in galaxies with active galactic nuclei (AGN) is
mainly due to reprocessed radiation, coming from the AGN UV-continuum
heated dust.  This radiation has a black body spectrum with
temperatures typically between 100 and 1000\,K. \citet{ocn05} report
an ultraviolet light outburst in the nucleus of NGC\,1399 with a
maximal far-UV luminosity of $1.2\times 10^{39}$ erg\,s$^{-1}$ reached
in January 1999. Thus they claim NGC\,1399 to harbour a low-luminosity
active galactic nucleus (LLAGN).  We tested this hypothesis by
subtracting black body spectra with different temperatures in the
above mentioned range from our innermost spectra with the goal to
match the line strengths seen at radii $>$ 0\farcs2. The results from
this test were unfavourable for a scenario in which the LLAGN induced
non-stellar continuum causes the observed line strength changes, since
we were unable to model the index changes for both \na\/ and \co\/
simultaneously with one single black body spectrum of a given
temperature. We can go even further and force the indices to drop by
the right amount by subtracting a continuum with a very steep slope
(in comparison with the already mentioned black body spectra).
However, when we use this steeper slope to match the index drops, the
effective velocity dispersion measured from such a spectrum
($\sigma$\,=\,384\,\kms) does not agree well with our results from
Sect.~\ref{sec:kinematics}. Moreover, the resulting spectrum does not
resemble the observed central galaxy spectrum (i.e. different
continuum shape and slope), which it is supposed to model.  Therefore,
we conclude that the presence of an AGN-like continuum is not very
likely to be the reason for the central line strength and velocity
dispersion changes observed in NGC\,1399.

Coming back to the stellar population issue it is very interesting to
note that the \citet{geb07} imaging data of NGC\,1399 (HST/WFPC2 PC
chip) show a blue nucleus ($r\,\le\,0\farcs1$) with weak evidence for
a central light excess at the same radii (see
Fig.\ref{fig:1399lum}). Their (B\,--\,V) colour drops by about 0.1 mag
in comparison with the surrounding region, which appears to be
constant in colour (their Figure 1, bottom panel). Recent HST/ACS
observations of NGC\,1399 in $g$ and $z$ bands \citep[][also Andres
Jord{\'a}n, private communication]{cote07} show very weak evidence for
a central light excess and also very mild evidence for a bluer colour,
but exhibit prominent PA and ellipticity changes within $r \le
0\farcs2$ consistent with the HST/WFPC2 data. We note, that the core
region of NGC\,1399 with a break radius of about 2\arcsec\/ (see
Fig.~\ref{fig:1399lum}) is much larger than the scales discussed here.

The information from the imaging, combined with the observed $\sigma$
and line strength drops at similar radii provide good evidence for the
scenario of a central dynamically cold stellar subsystem with
different stellar populations as compared to the surrounding galaxy.

Until stellar population models in the near-IR become available we can
only perform a limited stellar population analysis by using the
empirical relations found in the \citet{silva2008} study (their Figure
14). For early-type galaxies with old stellar populations they found a
good linear correlation between the \na\/ index and the optical
metallicity indicator [MgFe]$^\prime$ as well as \co\/ and
[MgFe]$^\prime$. If the central index drops are caused by a
metallicity change alone (i.e. lower metallicity), we would expect to
see near-IR index changes consistent with the relations seen in
\citet{silva2008}. However, this is not the case. The index change
seen for \na\/ is much too weak compared to \co, or vice
versa. \citet{silva2008} show also that centres of Fornax ellipticals
with optical signatures of a younger population (i.e. strong H$\beta$,
indicating luminosity weighted ages of around 2\,Gyr) have a stronger
\na\/ index in comparison with old galaxies at the same central
velocity dispersions. Also \co\/ is expected to increase for
intermediate ages ($\sim$\,1-2\,Gyr) since the fraction of thermally
pulsating asymptotic giant branch stars near or above the tip of the
first ascent giant branch goes up \citep[][and references
therein]{maraston05}.  A scenario which is then more consistent with
the empirical relations from \citet{silva2008} is a drop in
metallicity combined with a younger age where the \na\/ index reacts
stronger to the age change than the \co\/ index. Alternatively, one
can consider -- in addition to a drop in metallicity, abundance ratio
differences between the main body of NGC\,1399 and the proposed
central stellar subsystem, which would explain the relative changes of
\co\/ and \na.  While a combination of metallicity, age and abundance
ratio changes are an attractive explanation for the observed
line strength maps, a confirmation of the above scenarios clearly
awaits the more detailed predictions of stellar population models.

\section{Concluding remarks}

In this paper we present for first time high spatial resolution K-band
maps for the kinematical and near-IR spectral properties of the centre
of the giant cD galaxy in the Fornax cluster, NGC\,1399. We confirm
the presence of a central velocity dispersion dip within $r \le
0\farcs2$ previously seen in the long-slit studies of \citet{h06} and
\citet{geb07}. Our velocity dispersion map gives evidence for a
non-symmetric structure towards the centre by showing three $\sigma$
peaks to the north-east, south-east and west of the galaxy
centre. Such a complex structure in the velocity dispersion has not
been seen earlier in other early-type galaxies. The west -- south-east
peaks seem to be symmetric about the galaxy centre, while the
north-east peak is essentially off-centred, nevertheless within our
error analysis we consider it to be real. We note that the nuclear
region of NGC\,1399 does not seem to be obscured by dust (Andres
Jord{\'a}n, private communication).

Additionally we measure two near-IR line strength indices at
unprecedented spatial resolution.  The most important features we
observe in our 2-dimensional line strength maps are drops in \na\/ and
\co\/ line strength in the nuclear region of the galaxy, coinciding
spatially with the drop in $\sigma$.

As an explanation for the line strength and velocity dispersion drops
observed in the central region we discuss a scenario where NGC\,1399
harbours a dynamically cold subsystem with a distinct stellar
population.  Dynamically this can be realised by a central stellar
disc or a globular cluster (even a dwarf galaxy) having fallen into
the centre of NGC\,1399 on a purely radial orbit \citep{geb07}. The
near-IR index changes observed in our data suggest that the spectral
properties of this subsystem are directed by some mixture of
metallicity, age and abundance ratios effects. More precise
constraints on the nature of the stellar population within this cold
component awaits more detailed stellar population models.

\acknowledgements{We would like to thank the Garching and Paranal
astronomers who provided support during the SINFONI Science
Verification runs. Special thanks go to Ryan Houghton for valuable
help on preparing the observations and providing us with the NGC\,1399
NACO luminosity profile. We are especially grateful to Michele
Cappellari for supporting our project with additional velocity
template stars. The comparison between different kinematics studies in
NGC\,1399 could not have been possible without assistance from Roberto
Saglia and Karl Gebhardt. We also would like to thank Andrea
Modigliani for valuable help with the SINFONI pipeline and Karina
Kj{\ae}r for multiple discussions on 3D spectroscopy data
reduction. Finally, we thank the anonymous referee for his/her
important suggestions which certainly made this paper more complete.}

\bibliographystyle{aa}
\bibliography{n1399_sinfo.bib}

\end{document}